\documentclass{llncs}

\usepackage{url}

\usepackage{amssymb}
\usepackage{amsmath}
\usepackage{bm}
\usepackage{latexsym}
\usepackage{epsfig}
\usepackage{hyperref}
\usepackage{alltt}
\usepackage{wasysym}
\usepackage[ruled,vlined,linesnumbered]{algorithm2e}
\usepackage{framed}
\usepackage{paralist}
\usepackage{comment}

\usepackage{graphicx}
\usepackage{subfigure}
\usepackage[pdf]{pstricks}
\usepackage{pstricks-add}
\usepackage{pdfpages}
\usepackage{titlepic}

\usepackage[ruled,vlined,linesnumbered]{algorithm2e}
\DontPrintSemicolon

\usepackage{nzproof,theorem}

\begin{document}

%+Title
\title{Selection in the Presence of Memory Faults, with Applications to In-place Resilient Sorting}
\author{
    Tsvi Kopelowitz
    \and
    Nimrod Talmon}
\institute{
    Weizmann Institute of Science, Rehovot, Israel.\\ \email{\{kopelot,nimrodtalmon77\}@gmail.com}}
\maketitle
%-Title

\begin{abstract}

The selection problem, where one wishes to locate the $k^{th}$ smallest element in an unsorted array of size $n$, is one of the basic problems studied in computer science. The main focus of this work is designing algorithms for solving the selection problem in the presence of memory faults. These can happen as the result of cosmic rays, alpha particles, or hardware failures.

Specifically, the computational model assumed here is a faulty variant of the RAM model (abbreviated as \emph{FRAM}), which was introduced by Finocchi and Italiano~\cite{resilient_data_structures}. In this model, the content of memory cells might get corrupted adversarially during the execution, and the algorithm cannot distinguish between corrupted cells and uncorrupted cells. The model assumes a constant number of reliable memory cells that never become corrupted, and an upper bound $\delta$ on the number of corruptions that may occur, which is given as an auxiliary input to the algorithm. An output element is correct if it has rank between $k-\alpha$ and $k+\alpha$ in the input array, where $\alpha$ is the number of corruptions that occurred during the execution of the algorithm. An algorithm is called \emph{resilient} if it always outputs a correct answer.

The main contribution of this work is a deterministic resilient selection algorithm with optimal $O(n)$ worst-case running time. Interestingly, the running time does not depend on the number of faults, and the algorithm does not need to know $\delta$. As part of the solution, several techniques that allow to sometimes use non-tail recursion algorithms in the FRAM model are developed. Notice that using recursive algorithms in this model is problematic, as the stack might be too large to fit in reliable memory.

The aforementioned resilient selection algorithm can be used to improve the complexity bounds for resilient $k$-d trees developed by Gieseke, Moruz and Vahrenhold~\cite{kd_trees}. Specifically, the time complexity for constructing a $k$-d tree is improved from $O(n\log^2 n + \delta^2)$ to $O(n \log n)$.

Besides the deterministic algorithm, a randomized resilient selection algorithm is developed, which is simpler than the deterministic one, and has $O(n + \alpha)$ expected time complexity and $O(1)$ space complexity (i.e., is in-place). This algorithm is used to develop the first resilient sorting algorithm that is in-place and achieves optimal $O(n\log n + \alpha\delta)$ expected running time.
\end{abstract}

\section{Introduction}
\label{sec:introduction}

Computing devices are becoming smaller and faster. As a result, the likelihood of soft memory errors (which are not caused by permanent failures) is increased. In fact, a recent practical survey~\cite{soft_error_rate} concludes that a few thousands of soft errors per billion hours per megabit is fairly typical, which would imply roughly one soft error every five hours on a modern PC with 24 gigabytes of memory~\cite{CDK11}. The causes of these soft errors vary and include cosmic rays~\cite{radiation}, alpha particles~\cite{alpha_particles}, or hardware failures~\cite{hardware_errors}.

\subsection{The Faulty RAM Model}

To deal with these faults, the faulty RAM (FRAM) model has been proposed by Finocchi and Italiano~\cite{resilient_data_structures}, and has received some attention \linebreak \cite{optimal_resilient_dictionaries,resilient_and_external_memory,resilient_counters,dynamic_programming,resilient_sorting,resilient_dictionaries,kd_trees,resilient_priority_queues}. \linebreak In this model, an upper bound on the number of corruptions is given to the algorithm, and is denoted by $\delta$, while the \emph{actual} number of faults is denoted by $\alpha$ ($\alpha \leq \delta)$. Memory cells may become corrupted at any time during an algorithm's execution and the algorithm cannot distinguish between corrupted cells and uncorrupted cells. The same memory cell may become corrupted multiple times during a single execution of an algorithm. In addition, the model assumes the existence of $O(1)$ reliable memory cells, which are needed, for example, to reliably store the code itself. A cell is assumed to contain $\Theta(\log n)$ bits, where $n$ is the size of the input, as is usual in the RAM model.

One of the interesting aspects of developing algorithms in the FRAM model is that the notion of correctness is not always clear. Usually, correctness is defined with respect to the subset of uncorrupted memory cells and in a worst-case sense, implying that for an algorithm to be correct, it must be correct in the presence of any faulty environment, including an adversarial environment. For example, in the sorting problem the goal is to order the input elements such that the uncorrupted subset of the array is guaranteed to be sorted~\cite{resilient_data_structures}. In the FRAM model, an algorithm that is always correct (which is problem dependent) is called \emph{resilient}.

A naive way of implementing a resilient algorithm is by storing $2\delta+1$ copies of every piece of data. Writing is done by writing the same value to all copies, and reading is done by computing the majority of the copies. Using this technique, most if not all\footnote{The reason this might not be true is because it could depend on the correctness of the problem under the FRAM model. For example, the goal of finding the \emph{exact} $k$-order statistics is not achievable  in this model, as is explained in Section~\ref{sec:preliminaries}.} non-resilient algorithms can be made resilient with $O(\delta)$ multiplicative overhead in time and space complexity.

\subsection{Previous Work}

A summary of the algorithms and data structures that have been developed in the FRAM model is given next.

\paragraph{Resilient Searching:} 
		
		Finocchi and Italiano~\cite{resilient_data_structures} and
		Finocchi, Grandoni and Italiano~\cite{resilient_sorting},
		developed an almost optimal resilient searching algorithm,
		which finds an element in a sorted array of size $n$ in $O(\log n + \delta^{1+\epsilon})$,
		where uncorrupted elements are guaranteed to be found.
		The main idea is to perform a slow reliable verification step once in every $O(\delta)$
		fast but unreliable binary search steps.    
		A somewhat natural lower bound of $O(\log n + \delta)$ was proven there as well.
		A matching upper bound was developed by
		Brodal, Fagerberg, Finocchi, Grandoni, Italiano, J{\o}rgensen, Moruz and M{\o}lhave~\cite{optimal_resilient_dictionaries},
		using a different method.

\paragraph{Resilient Dictionaries:}

	The dynamic counterpart of searching is the dictionary data structure.
	An optimal resilient dictionary,
	supporting updates (insertions and deletions) and queries (searches)
	in $O(\log n + \delta)$ amortized time per operation,
	was developed by Brodal et al.~\cite{optimal_resilient_dictionaries}.
	Again, uncorrupted elements are guaranteed to be found.

\paragraph{Resilient Sorting:}

	Finocchi et al.~\cite{resilient_data_structures,resilient_sorting},
	developed a resilient sorting algorithm,
	sorting an array of size $n$ in $O(n\log n + \alpha\delta)$ time.
	The uncorrupted subset of the array is guaranteed to be sorted.
	The algorithm is an iterative version of Mergesort,
	with a resilient merging step.
	A matching (and somewhat surprising) lower bound
	was proven there as well.
	
\paragraph{Resilient Priority Queues:}

	Another basic data structure, a resilient priority queue,
	was developed by J{\o}rgensen, Moruz and M{\o}lhave~\cite{resilient_priority_queues}.
	The data structure supports \emph{insert} and \emph{deletemin} in $O(\log n + \delta)$ amortized time,
	where the \emph{deletemin} operation returns either the minimum element among the uncorrupted elements,
	or a corrupted element.
	A matching lower bound was given there as well.	
		
\paragraph{Resilient Counters:}

	Brodal, Gr{\o}nlund, J{\o}rgensen, Moruz and \linebreak M{\o}lhave~\cite{resilient_counters},
	developed several resilient counters,
	supporting increments and queries,
	where the result of a query is an $\alpha$-additive approximation
	to the number of increments performed until the query.
	While the proven lower bound of $\Omega(\delta)$ space and time
	is not achieved, several interesting tradeoffs are presented there.

\paragraph{Dynamic Programming:}	
	
	Caminiti, Finocchi and Fusco~\cite{first_dynamic_programming}
	and Caminiti, Finocchi, Fusco and Silvestri~\cite{dynamic_programming},
	developed a resilient and cache-oblivious dynamic programming meta algorithm,
	computing the correct answer with high probability,
	using $O(n^d + \delta^{d+1})$ and $O(n^d + n\delta)$ space,
	where $d$ is the dimension of the table of the dynamic programming.

\paragraph{Resilient External Memory Algorithms:}

	The problem of designing algorithms that are simultaneously
	cache efficient and resilient was addressed by
	Brodal, J{\o}rgensen, Gr{\o}nlund and M{\o}lhave~\cite{resilient_and_external_memory}.
	They showed matching upper \linebreak bounds and lower bounds
	for a deterministic and randomized dictionary,
	a deterministic priority queue,
	and a deterministic sorting algorithm.

\paragraph{$k$-d Trees:}

	The problem of $k$-means clustering in the presence of memory faults
	was addressed by Gieseke et al.~\cite{kd_trees}.
	They developed a resilient $k$-d tree,
	supporting orthogonal range queries in
	$O(\sqrt{n\delta} + t)$ where $t$ is the is the size of the output.

\subsection{Results}

\paragraph{Deterministic Resilient Selection Algorithm:}

The main focus of this work is on the \emph{selection problem} (sometimes called the $k$-order statistic problem) in the FRAM model, where one wishes to locate the $k^{th}$ smallest element in an unsorted array of size $n$, in the presence of memory faults. The following main theorem is proved in Section~\ref{sec:resilient_deterministic_selection_algorithm}.

\begin{theorem}
There exists a deterministic resilient selection algorithm with time complexity $O(n)$.
\end{theorem}

Interestingly, the running time does not depend on the number of faults. Moreover, the algorithm does not need to know $\delta$ explicitly. The selection problem is a classic problem in computer science. Along with searching and sorting, it is one of the basic problems studied in the field, taught already at undergraduate level (e.g.,~\cite{CLRS}). The $k$-order statistic of a set of samples is a basic concept in statistics as well (e.g.,~\cite{order_statistics}). There are numerous applications for the selection problem, thus devising efficient algorithms is of practical interest. The textbook algorithm by Blum, Floyd, Pratt, Rivest and Tarjan~\cite{median_of_medians_algorithm}
achieves linear time complexity in the (non-faulty) RAM model.

When considering the selection problem in the FRAM model, the first difficulty is to define correctness\footnote{The common notion of considering only the non-corrupted elements is somewhat misleading in the selection problem. This is because of the difficulty of not being able to distinguish between corrupted and uncorrupted data.}.  To this end, the correctness definition used here allows to return an element, which may even be corrupted, whose rank is between $k-\alpha$ to $k+\alpha$ in the input array. Notice that when $\alpha = 0$ this definition coincides with the non-faulty definition (for a formal definition see Section~\ref{sec:preliminaries}).

\paragraph{Randomized Resilient Selection Algorithm:}

Besides the deterministic algorithm, a randomized and in-place counterpart is developed as well.
A randomized algorithm in the FRAM model is an algorithm that can use random coins. The faults are still adversarial, but the adversary cannot see the random coins of the algorithm, and the algorithm must be correct with probability $1$, regardless of the coin tosses. The randomized selection algorithm is simpler than to the deterministic one, and is likely to beat the deterministic algorithm in practice.
The following theorem is proven in Section~\ref{sec:resilient_randomized_selection_algorithm}.

\begin{theorem}
There exists a randomized in-place resilient selection algorithm with expected time complexity $O(n+\alpha)$.
\end{theorem}

\paragraph{Resilient $k$-d Trees:}

The selection algorithm presented here can be used to improve the complexity bounds for resilient $k$-d trees developed by Gieseke et al.~\cite{kd_trees}. There, a deterministic resilient algorithm for constructing a $k$-d tree with $O(n\log^2 n + \delta^2)$ time complexity is shown. This can be improved to $O(n \log n)$ by using the deterministic resilient selection algorithm developed here.

\begin{theorem}
There exists a resilient $k$-d tree which can be constructed in deterministic $O(n \log n)$ time.
It supports resilient orthogonal range queries in $O(\sqrt{n \delta} + t)$ time for reporting $t$ points.
\end{theorem}

\paragraph{Resilient Quicksort Algorithms:}

The problem of sorting in the FRAM model is also revisited, as an application of the resilient selection algorithm. Finocchi et al.~\cite{resilient_sorting}, already developed a resilient Mergesort algorithm, sorting an array of size $n$ in $O(n\log n + \alpha\delta)$ time, where the uncorrupted subset of the array is guaranteed to be sorted. They also proved that this bound is tight. In Section~\ref{sec:resilient_quicksort_algorithms}, a new in-place randomized sorting algorithm which resembles Quicksort and runs in $O(n\log n + \alpha\delta)$ expected time is presented. This sorting algorithm uses the randomized selection algorithm as a black box.
The following theorem is proven in Section~\ref{sec:resilient_quicksort_algorithms}.

\begin{theorem}
There exists a resilient deterministic sorting algorithm with worst-case running time of $O(n\log n+\alpha\delta)$, and a resilient randomized in-place sorting algorithm with expected running time of $O(n\log n+\alpha\delta)$.
\end{theorem}

\subsection{Recursion}

In the (non-faulty) RAM model the recursion stack needs to reliably store the local variables, as well as the frame pointer and the program counter. Corruptions of this data can cause the algorithm to behave unexpectedly, and in general the recursion stack cannot fit in reliable memory. Some new techniques for implementing a specific recursion stack which suffices for solving the selection problem are developed in Section~\ref{sec:recursion_implementation}. These techniques are used to develop the resilient deterministic selection algorithm presented in Section~\ref{sec:resilient_deterministic_selection_algorithm}. It is likely that these techniques can be used to help implement recursive algorithms for other problems in the FRAM model. The main technique developed here which allows to use non-tail recursion in the FRAM model is somewhat general, and can be used due to the following four points:

\begin{enumerate} 

\item \emph{Easily Inverted Size Function:} When performing a recursive call, the function which determines the size of the input to the recursive call can be easily inverted, while needing only $O(1)$ bits to maintain the data needed to perform the inversion.

\item \emph{Small Depth:} The depth of the recursion is bounded by $O(\log n)$ and so using $O(1)$ bits per level can fit in reliable memory.

\item \emph{Verification:} A linear verification procedure is used such that once a recursive call finishes, if the procedure accepts, then the algorithm may proceed even if some errors did occur in the recursive call. The main point here is that even though errors occurred, continuing onwards does not hurt the correctness.

\item \emph{Amortization:} If the verification procedure fails, then the number of errors which caused the failure is linear in the amount of time spent on the recursive call (not counting other verification procedures that failed within it). This means that the amortized cost of each corruption is $O(1)$.

\end{enumerate}

The only previous work done in the FRAM model for non-tail recursion was done by Caminiti et al.~\cite{dynamic_programming} where they developed a recursive algorithm for solving dynamic programming. However, the recursion inherited in the problem of dynamic programming is simpler compared to the recursion treated in the selection problem, due to the structural behavior of the dynamic programming table (the recursions depend on positioning within the table, and not on the actual data). Moreover, their solution only works with high probability (due to using fingerprints for the verification procedure).

\subsection{Related Work}

Other models and techniques to deal with memory corruptions do exist.
Some of them are given here, with an emphasis on their relation to the FRAM model.

\paragraph{Error Correcting Codes:}
		
		The field of error correcting codes and error detecting codes
		deals with the problem of reliably
		transmitting a message over a faulty communication channel.
		This is achieved by adding redundancy to the message (e.g., checksums).
		For a survey, see, e.g., ~\cite{ecc_survey}.
		The solutions developed in this field do not treat the implications of corruptions
		to the computation performed on the data.
		Therefore, applying these methods to the FRAM model
		in a non-naive way is not trivial.   
		
\paragraph{Error Correcting Memory:}
		
		Error detecting and correcting codes can be 
		implemented in the hardware itself (e.g., \cite{error_correcting_memory}).
		While this solution has its advantages,
		it imposes some costs in performance and money.
		
\paragraph{Pointer-Based Data Structures:}

		Aumann and Bender~\cite{aumann_bender} addressed the problem of losing
		data in a pointer-based data structures due to pointer corruptions.
		The data structures suggested by them incur only a small overhead in space and time,
		and guarantee an upper bound on the amount of uncorrupted data that can be lost due
		to pointer corruptions.
		This is in contrast to the FRAM model, where no uncorrupted  data is allowed to be lost.
		
\paragraph{Fault-Tolerant Parallel and Distributed Computation:}
	
		Extensive research on fault tolerance have been done in the field of 
		parallel and distributed computation.
		For a survey, see~\cite{fault_tolerant_distributed_computation}.
		The work done in this field deals with resiliency with respect to
		faulty processors or communication links, in contrast to the faulty memory
		which is assumed in the FRAM model.
		Some of the work assume the existence of fault detection
		hardware, therefore allowing the system to distinguish between faulty and non-faulty data,
		differently from the FRAM model.

\paragraph{Checkers:}

		Blum, Evans, Gemmell, Kannan and Naor~\cite{checkers} 
		addressed the problem of checking memory correctness in the presence of faults.
		In this model, the data structure is viewed as being controlled by an adversary.
		The goal of the checker,
		which is allowed to use a small amount of reliable memory,
		is to detect every deviation from the expected data, 
		with high probability. In the FRAM model, the goal is not to detect the 
		memory corruptions, but instead, to always behave correctly on the uncorrupted subset of the data.

\paragraph{Fault-Tolerant Sorting Networks:}
		
		Fault tolerance have been investigated in the context of sorting networks.
		Assaf and Upfal~\cite{fault_tolerant_sorting_networks} developed a resilient sorting network,
		with an $O(\log n)$ multiplicative overhead in the size of the network.
		The computational model is a sorting network and not a general purpose machine, as in the FRAM model.
				 
\paragraph{The Liar Model:} 

		In this model, the algorithm can access the data only through a noisy oracle.
		The algorithm queries the oracle and can possibly get a faulty answer (i.e., a lie).
		An upper bound on the number of these lies or a probability of a lie is assumed.
		See, e.g., ~\cite{liar_1} and~\cite{liar_2}.
		The data itself cannot get corrupted, therefore, 
		in this model, query replication strategies can be exploited,
		in contrast to the FRAM model.

\paragraph{Other Noisy Computational Model:}
		
		Several other noisy computational models have been investigated.
		Sherstov~\cite{robust_polynomials}, showed an optimal (in terms of degree) approximation polynomial
		that is robust to noise. Gacs and Gal~\cite{robust_circuits}, proved a lower bound on the number of gates
		in a noise resistant circuit.
		These works, as well as others, have more computational complexity theory flavour than the FRAM model,
		and treat different computational models from the FRAM model.

\subsection{Organization}

The paper is organized as follows.
In Section~\ref{sec:preliminaries} some definitions and preliminaries are given.
In Section~\ref{sec:resilient_randomized_selection_algorithm} the randomized selection algorithm is discussed,
followed by a discussion of the deterministic selection algorithm, in Section~\ref{sec:resilient_deterministic_selection_algorithm}.
The discussion of the stack and recursion implementation is treated independently and deferred to Section~\ref{sec:recursion_implementation}.
A discussion on the application of the resilient selection algorithm to resilient $k$-d trees
is in Section~\ref{sec:resilient_kd_trees}. 
Finally, the in-place quicksort sorting algorithm is shown in Section~\ref{sec:resilient_quicksort_algorithms}.

\section{Preliminaries}
\label{sec:preliminaries}

\subsection{Definitions}

Let $X$ be an array of size $n$ of elements taken from a totally ordered set. Let $X^0$ denote the state of $X$ at the beginning of the execution of an algorithm $A$ executed on $X$. Let $\alpha \leq \delta$ be the number of corruptions that occurred during such execution.

\begin{definition}
Let $X$ be an array and let $e$ be an element. The \emph{rank of $e$ in $X$} is defined as $rank_X(e) = |\{i : X[i] \leq e\}|$. The \emph{$\alpha$-rank of $k$ in $X$} is defined as $\alpha\text{-}rank_X(k) = \{e : rank_X(e) \in [k - \alpha, k + \alpha]\}$.
\end{definition}

Notice that the $\alpha$-rank of $k$ in $X$ is an interval containing the elements whose rank in $X$ is not smaller than $k - \alpha$ and not larger than $k + \alpha$. In particular, if $\alpha \geq n$, this interval is equal to $[-\infty,\infty]$.
Moreover, if $\alpha = 0$, then this interval is equal to the $k$-order statistic, thus coincides with the non-faulty definition.

\begin{definition}
\label{def:resilient-selection-algorithm}
A \emph{resilient $k$-selection algorithm} is an algorithm that is given an array $X$ of size $n$ and an integer $k$, and returns an element $e \in \alpha\text{-rank}_{X^0}(k)$, where $\alpha \leq \delta$ is the number of faults that occurred during the execution of the algorithm.~\end{definition}

Notice that if $\alpha=0$, then this definition coincides with the common non-faulty definition. That is, if no faults occur during an execution of a resilient selection algorithm, it should locate the exact $k$-order statistic.
Moreover, if $\alpha>0$, no algorithm can return the \emph{exact} $k$-order statistics, due to corruptions that can happen at the beginning of the execution.
Notice also that because the algorithm cannot distinguish between corrupted and uncorrupted memory cells, it may return an element which was not present in the array at the beginning of the execution.

\subsection{Basic Procedures}

\begin{lemma}
\label{def:ranking-procedure}
There exists a \emph{resilient ranking procedure} with time complexity $O(n)$, that is given an array $X$ of size $n$ and an element $e$, and returns an integer $k$ such that $e \in \alpha\text{-rank}_{X^0}(k)$.
\end{lemma}

\begin{pf}
A resilient ranking procedure can be implemented by scanning $X$ while counting the number of elements smaller or equal to $e$, denoted by $k$. If $\alpha = 0$, then $k = rank_X(e)$. If $\alpha > 0$, then $e \in \alpha\text{-rank}_{X^0}(k)$, because each corruption can change at most one memory cell, changing the rank of $e$ in $X$ by \linebreak at most $1$.
\end{pf}

\begin{lemma}
\label{resilient_partition_algorithm}
There exists a \emph{resilient partition procedure} with time complexity $O(n)$ and space complexity $O(1)$, that is given an array $X$ of size $n$ and an element $e$, and reorders $X$ such that the uncorrupted elements smaller (larger) than $e$ are placed before (after) $e$, and returns an element $k$ such that $e \in \alpha\text{-rank}_{X^0}(k)$.
\end{lemma}

\begin{pf}
A resilient partition procedure can be implemented by scanning $X$ while counting the number of elements smaller or equal to $e$, denoted by $k$, such that whenever an element smaller than $e$ is encountered it is swapped with the element at position $k + 1$.
\end{pf}

Notice that both procedures compute an integer $k$ such that $e \in \alpha\text{-rank}_{X^0}(k)$. Let $rank^c_X(e)$ denote the value $k$ computed by either procedure, such that whenever the notation $rank^c_X(e)$ will be used, it will be understood from the context which procedure is used. Notice that if $\alpha = 0$, then $rank^c_X(e) = rank_X(e)$.

\section{Randomized Resilient Selection Algorithm}
\label{sec:resilient_randomized_selection_algorithm}

As a starter, consider the following randomized resilient selection algorithm, denoted by \emph{Randomized-Select}. The algorithm is an adaptation of the randomized non-resilient selection algorithm by Hoare~\cite{hoare_selection}, with the following modification. The algorithm maintains an interval $[lb,ub]$, where $lb$ ($ub$) is a lower (upper) bound. When the algorithm queries the array $X$ at index $i$, the value $x$ is chosen to be $x = min(max(X[i], lb), ub)$. This guarantees that even a faulty value is within the bounds.

All variables (i.e., $l$, $r$, $lb$, $ub$, $x_p$, $p$, $k$) are stored using reliable memory cells.

\vspace{15px}
\begin{algorithm}[H]
\SetKwFor{Loop}{repeat}{}{}

$l \leftarrow 1$, $r \leftarrow n$, $lb \leftarrow -\infty$, $ub \leftarrow \infty$\;
\Loop{}{
	$x_p \leftarrow$ random element from $X[l,r]$\;
	$x_p \leftarrow min(max(x_p,lb),ub)$\;
	partition $X$ around $x_p$ \# using the algorithm from Lemma~\ref{resilient_partition_algorithm}\;
	\# Let $p$ denote $rank^c_X(x_p)$\;
	\uIf{$p = k$}
	{\Return $x_p$\;}
	\uElseIf{$p > k$}
	{$r \leftarrow p - 1$, $ub \leftarrow x_p$\;}
	\ElseIf{$p < k$}
	{$k \leftarrow k - p$, $l \leftarrow p + 1$, $lb \leftarrow x_p$\;}
}

\caption{Randomized-Select($X$, $k$)}
\end{algorithm}
\vspace{15px}

\begin{theorem}\label{theorem:random_select}
There exists a randomized in-place resilient selection algorithm with expected time complexity $O(n+\alpha)$.
\end{theorem}

\begin{pf}
Correctness is proven by induction on the size of the array. The base case of size $1$ is obvious. For the induction step, assume that for arrays of size smaller than $n$ the algorithm returns an element $e \in \alpha\text{-}rank_{X^0}(k)$. Consider an execution of the algorithm on an array of size $n$. Let $\alpha_1$ denote the number of corruptions that occurred during the first iteration, and let $\alpha'$ denote the number of corruptions that occurred during the rest of the execution ($\alpha = \alpha_1 + \alpha'$). During the first iteration of the algorithm, if $p = k$, then $e = x_p$ is returned and correctness follows from the definition of the resilient partition procedure, and from the fact that $x_p$ is maintained in reliable memory. Otherwise, assume without loss of generality, that $p < k$. The case where $p > k$ is symmetric.

The second iteration considers a sub-array $X' = X[p + 1, r]$ of size $n' < n$. Therefore, by the induction hypothesis, $e \in \alpha'\text{-}rank_{X'}(k)$. It is guaranteed that $e \geq x_p$, because $e$ is taken to be $min(max(e,lb),ub)$. Therefore, $e$ is larger than all the uncorrupted elements in $X[1:p]$. Each corruption that occurred during the first iteration can change the rank of $e$ by at most $1$, therefore \linebreak $e \in (\alpha' + \alpha_1)\text{-}rank_X(k) = \alpha\text{-}rank_X(k)$. Notice that the above proves that the algorithm is correct with probability $1$.

With regard to the expected time complexity, let $t$ denote the number of iterations the algorithm does. If there are no faults (i.e., $\alpha = 0$), then the probability of choosing a pivot $x_p$ such that $rank_X(x_p) \in [\frac{n}{4}, \frac{3n}{4}]$ is $\frac{1}{2}$. However, there are two types of possible corruptions. The first type is corruptions of elements that are used as pivot elements. The second type is corruptions of elements that are not used as pivot elements. Let $\alpha'$ ($\alpha''$) denote the number of corruptions of the first (second) type.

Consider corruptions of the first type. Let $i_0,\ldots,i_t$ be indices of iterations such that $i_0$ is the first iteration, $i_t$ is the last iteration, and for every $j > 0$, $i_{j+1}$ is the first iteration after $i_j$ such that $n_{i_{j+1}} \leq \frac{3}{4}n_{i_{j}} + \alpha_{i_j}$, where $n_{i_j}$ denotes the size of the sub-array at the beginning of the $i_j^{th}$ iteration and $\alpha_{i_j}$ denotes the number of corruptions that occurred between the $i_j^{th}$ iteration and the $(i_{j+1} - 1)^{th}$ iteration. It follows that $\sum_{j = 0}^{t - 1} \alpha_{i_j} = \alpha'$ and $n_{i_j} \leq \left( \frac{3}{4} \right)^j n + \sum_{k=0}^{j-1} \left( \frac{3}{4} \right)^{j-k-1} \alpha_{i_k}$. Let $Y_j$ denote the number of iterations between the $i_j^{th}$ iteration and the $(i_{j+1} - 1)^{th}$ iteration (i.e., $Y_j = i_{j+1} - i_j$). $Y_j$ is a random variable with a geometric distribution, and $\mathbb{E}(Y_j) \leq 2$, by a similar reasoning as in the non-faulty case. Notice that $\mathbb{E}(Y_j) \leq 2$ even when conditioned on earlier iterations. It follows that if there are only corruptions of the second type, then the running time is bounded by $\sum_{j=0}^{t} O(n_{i_j}) Y_j$.

Consider corruptions of the second type. For a sub-array of size $n'$, the probability that the adversary corrupts the pivot element using $1$ corruption is $1/n'$, because the adversary cannot see the random coins used by the algorithm. A corrupted pivot can result in up to $O(n')$ extra work. Therefore, the expected cost of a corrupted pivot is $O(1)$. To conclude, the expected time complexity is as follows\footnote{For simplicity, for $j > t$, $Y_j$ and $n_{i_j}$ are defined to be $0$.}:

\begin{align*}
\mathbb{E}\left[T(n)\right] &\leq \mathbb{E} \left[\sum_{j=0}^{t-1} n_{i_j} Y_j \right] + O(\alpha'') \leq \mathbb{E} \left[\sum_{j=0}^{\infty} n_{i_j} Y_j \right] + O(\alpha'')\\
&\leq \sum_{j=0}^{\infty} \mathbb{E} \left[n_{i_j} Y_j \right] + O(\alpha'')\\
&\leq \sum_{j=0}^{\infty} \mathbb{E} \left[ \left[ \left( \frac{3}{4} \right)^j n + \sum_{k=0}^{j-1} \left( \frac{3}{4} \right)^{j-k-1} \alpha_{i_k} \right] Y_j \right] + O(\alpha'')\\
&\leq \mathbb{E} \left[ Y_j \right] n \sum_{j=0}^{\infty} \left( \frac{3}{4} \right)^j + \sum_{j=0}^{\infty} \sum_{k=0}^{j-1} \mathbb{E} \left[ \left( \frac{3}{4} \right)^{j-k-1}    \alpha_{i_k} Y_j \right] + O(\alpha'')\\
\end{align*}

Notice that the expectation of $Y_j$ is at most $2$, even when conditioned on $\alpha_{i_k}$, for $k < j$ (i.e., $\mathbb{E}\left[ Y_j \right | \alpha_{i_k}] \leq 2$, for $k < j$). Therefore, using total expectation, it follows that, for $k < j$:
\begin{align*}
\mathbb{E}\left[\alpha_{i_k} Y_j \right] &= \sum_{\alpha_{i_k} = z} \mathbb{E}\left[Y_j \alpha_{i_k} | \alpha_{i_k} = z\right] \mathbb{P}\left[\alpha_{i_k} = z\right]\\
&= \sum_{\alpha_{i_k} = z} z \mathbb{E}\left[Y_j | \alpha_{i_k} = z\right] \mathbb{P}\left[\alpha_{i_k} = z\right]\\
&\leq 2 \sum_{\alpha_{i_k} = z} z \mathbb{P}\left[\alpha_{i_k} = z\right]\\
&= 2 \mathbb{E}\left[\alpha_{i_k}\right]
\end{align*}

Therefore,
\begin{align*}
\mathbb{E}\left[T(n)\right] &\leq 2 n \sum_{j=0}^{\infty} \left( \frac{3}{4} \right)^j + \sum_{j=0}^{\infty} \sum_{k=0}^{j - 1} 2 \mathbb{E} \left[ \left( \frac{3}{4} \right)^{j-k-1} \alpha_{i_k} \right] + O(\alpha'')\\
&\leq 8 n + 2 \sum_{j=0}^{\infty} \sum_{k=0}^{j - 1} \mathbb{E} \left[ \left( \frac{3}{4} \right)^{j-k-1} \alpha_{i_k} \right] + O(\alpha'')\\
&= O(n) + \sum_{j=0}^\infty O(\alpha_{i_j}) + O(\alpha'') = O(n) + O(\alpha') + O(\alpha'') = O(n + \alpha)
\end{align*}
\end{pf}

\section{Deterministic Resilient Selection Algorithm}
\label{sec:resilient_deterministic_selection_algorithm}

The following deterministic resilient selection algorithm is similar in nature to the non-resilient algorithm by Blum, Floyd, Pratt, Rivest, and Tarjan~\cite{median_of_medians_algorithm}, but several major modifications are introduced in order to make it resilient. The algorithm is presented in a recursive form, but the recursion is implemented in a very specific way, as explained in Section~\ref{sec:recursion_implementation}.

In the non-faulty RAM model, the recursion stack needs to reliably store the local variables, as well as the frame pointer and the program counter. Corruptions of this data can cause the algorithm to behave unexpectedly, and in general the recursion stack cannot fit in reliable memory. Therefore, a special recursion implementation is needed. 

Generally, a recursive computation can be thought of as a traversal on a recursion tree $T$, where the computation begins at the root. Each internal node $u \in T$ performs several recursive calls, which can be partitioned into two types: the \emph{first} type and the \emph{second} type. Each node performs at least one call of each type, and the calls may be interleaved. The idea is for each node $u$, to locate the $k_u^{th}$ smallest element in the array $X_u$ of size $n_u$. However, due to corruptions, this cannot be guaranteed, therefore a weaker guarantee is used, as explained later.

\subsection{Algorithm Description}

The root of the recursion tree is a call to \emph{Determinstic-Select($X$, $k$, $-\infty$, $\infty$)}.
The computation of an inner node $u$ has two phases.

\subsubsection{First phase}

The goal of the first phase is to find a \emph{good} pivot, specifically, a pivot whose rank is in the range $[f_u, n_u-f_u]$, where $f_u = \lfloor \frac{3n_u}{10} \rfloor - \lfloor \frac{n_u}{11} \rfloor - 6$.\footnote{The exact choice of $f_u$ (which is a function of $n_u$, the size of the node $u$) relates to the recursion implementation as explained in Section~\ref{sec:recursion_implementation}. The idea is to always partition the array at a predetermined ratio, in order to provide more structure to the recursion, and this is what allows for the recursion size function to be easily invertible, as mentioned in Section~\ref{sec:introduction}. Notice that the $\lfloor \frac{n_u}{11} \rfloor$ could be picked to be $\lfloor \epsilon \cdot n_u \rfloor$ for any constant $\epsilon < \frac{1}{10}$, because this is needed for the running time of the algorithm, as explained in the proof of Theorem~\ref{theorem:det_select_alpha}.} Finding a pivot is done by computing the median of each group of five consecutive elements in $X$, followed by a recursive call of the first type, to compute the median of these medians. The process is repeated until a \emph{good} pivot is found.

\subsubsection{Second phase}

The goal of the second phase is to find a \emph{good} element. Specifically, an element whose rank is in $[k_u \pm n_v]$ where $v$ is a second type child of $u$. This will be shown to be sufficient\footnote{The exact choice of $[k_u \pm n_v]$ relates to the proof by induction for the correctness of the algorithm. The idea is that as long as less then $n_v$ corruptions occurred during the computation of $v$, the rank of the element located by $v$ is guaranteed, by induction, to be in these bounds. See the proof of Lemma~\ref{lem:analysis_1} and the proof of Lemma~\ref{lem:second_phase_repetitions}.}. This is done by making a recursive call of the second type, which considers only the relevant sub-array with the updated order statistic. Notice that, unlike the non-faulty selection algorithm, here the appropriate sub-array might be padded with more elements, so that the size of the sub-array is $n_u-f_u$. This is important for the recursion implementation, as explained in Section~\ref{sec:recursion_implementation}. If the returned value from the recursive call is not in the accepted range, the entire computation of the node repeats, starting from the first phase. Once a \emph{good} element is found, it is returned to the caller.

\newpage

\vspace{10mm}

\begin{algorithm}[H]

\SetKw{Or}{or}
\SetKw{Goto}{goto}

\# The algorithm uses the recursion implementation from Lemma~\ref{lem:recursion_lemma} \;
\Repeat{
	$rank^c_X(e) \in [k \pm n_v]$ \# $v$ is a second type child of the node
}{	
	\# Let $f$ denote $\lfloor \frac{3n}{10} \rfloor - \lfloor \frac{n}{11} \rfloor - 6$\;
	\Begin(First Phase){
		\Repeat{
			$p \in [f, n - f]$
		}{
			$X_m \leftarrow []$\;
			\For{$i \in [1..\lceil n/5 \rceil]$}{$X_m[i] \leftarrow$ median of $X[5i, 5i + min(4, n - 5i)]$\;}
			$x_p \leftarrow$ Deterministic-Select($X_m$, $\lceil |X_m|/2 \rceil$, $lb$, $ub$)\;
			partition $X$ around $x_p$ \# using the algorithm from Lemma~\ref{resilient_partition_algorithm}\;
			\# Let $p$ denote $rank^c_X(x_p)$\;
		}
	}
	
	\Begin(Second Phase){
		\uIf{$p = k$}{\Return $e = min(max(x_p, lb), ub)$\;}
		\uElseIf{$p > k$}{$e \leftarrow$ Deterministic-Select($X[1, n-f]$, $k$, $lb$, $x_p$)\;}
		\ElseIf{$p < k$}{$e \leftarrow$ Deterministic-Select($X[f, n]$, $k-f$, $x_p$, $ub$)\;}
	}
}

\Return $e = min(max(e, lb), ub)$\;

\caption{Deterministic-Select($X$, $n$, $k$, $lb$, $ub$)}
\end{algorithm}

\vspace{10mm}

Let $\alpha_u$ be the number of corruptions that occurred in $u$'s sub-tree. Each node uses two boundary values $lb_u$ and $ub_u$ which are used similarly to the bounds used in the randomized resilient algorithm.

The recursive calls are made with the parameters $X_u$, $n_u$, $k_u$, $lb_u$, $ub_u$, and each recursive call returns an element $x$. In Section~\ref{sec:recursion_implementation}, a recursion implementation with the following properties is described.

\begin{lemma}
\label{lem:recursion_lemma}
There exists a recursion implementation for the resilient deterministic selection algorithm with the following properties:
\begin{enumerate}
	\item
		The position of $X_u$, $n_u$, the return value, and program counter are reliable.\footnote{This means that these variables are correct,
		as long as no more than $\delta$ faults occurred.}
	\item 
		If $\alpha_u \leq n_u$, then $lb_u$, $ub_u$, $k_u$ are reliable.\footnote{This means that these variables are correct,
		as long as no more than $n_u$ faults occurred.}
	\item
		The time overhead induced by the implementation is $O(n_u)$ per call.
\end{enumerate}
\end{lemma}
The proof of the Lemma is given in Section~\ref{sec:recursion_implementation}.

\subsection{Analysis}

\begin{figure}
%\begin{center}
\psset{gridcolor=red,subgridcolor=blue}

\psset{unit=1.5cm}

\hspace*{-1.0in}
\begin{pspicture}(-2.5,0)(5,3)

\pscircle(3.5,3){0.5}
\rput(3.5,3){$u$}

\pscircle[linestyle=dashed](1,1.5){0.5}
\rput(1,1.5){$v_1$}
\pscircle(2.25,1.5){0.5}
\rput(2.25,1.5){$v_2$}
\pscircle[linestyle=dashed](3.5,1.5){0.5}
\rput(3.5,1.5){$v_3$}
\pscircle[linestyle=dashed](4.75,1.5){0.5}
\rput(4.75,1.5){$v_4$}
\pscircle(6,1.5){0.5}
\rput(6,1.5){$v_5$}

\psline(3.5,2.5)(1,2)
\psline(3.5,2.5)(2.25,2)
\psline(3.5,2.5)(3.5,2)
\psline(3.5,2.5)(4.75,2)
\psline(3.5,2.5)(6,2)

\psbrace[rot=90,ref=t,braceWidth=0.01](0.5,1)(2.75,1){$v_2$}
\psbrace[rot=90,ref=t,braceWidth=0.01](3,1)(4,1){$v_3$}

\end{pspicture}
%\end{center}

\caption{A node $u$ with five children: $v_1,\ldots,v_5$ is depicted. The nodes $v_1$,$v_3$,$v_4$ are first type children of $u$, while the nodes $v_2$,$v_5$ are second type children of $u$. The braces show the corruptions amortization, specifically, $v_2$ pays for $v_1$ and for itself and $v_3$ pays for itself.}
\label{fig:recursion_tree}
\vspace{0.2in}
\end{figure}
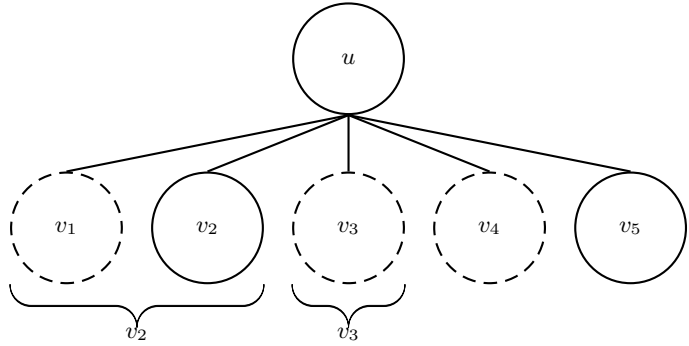

Let $u$ be a node. Let $V = (v_1,\ldots,v_{|V|})$ be $u$'s children. $v_1$ is always a first type node, and $v_{|V|}$ is always a second type node. Every second type child, except $v_{|V|}$, is followed by a first type child, therefore there cannot be two adjacent second type children (see Fig.~\ref{fig:recursion_tree}). Let $\alpha_u$ denote the number of corruptions that occur in $u$'s sub-tree and let $\alpha_u^{local}$ denote the number of corruptions that occur only in $u$'s data. Let $\alpha_u^{v_i}$ denote the number of corruptions that occur in $u$'s data between the execution of $v_i$ and the execution of $v_{i+1}$ (or until $u$ finishes its computation, if $v_i$ is the last child of $u$) and let $\alpha_v^0$ denote the number of corruptions that occur in $u$'s data before the execution of $v_1$. It follows that, $\alpha_u = \alpha_u^{local} + \sum_{v=0}^{|V|} \alpha_{v_i} = \sum_{v = 1}^{|V|} (\alpha_u^{v_i} + \alpha_{v_i})$. Let $X_u^0$ denote the state of $X_u$ at the beginning of $u$'s computation. Let $X_u^{v_i}$ denote the state of $X_u$ at the moment of the call to $v_i$.

\vspace{0.4in}

The following Lemmas are used to prove the correctness and the running time of \emph{Deterministic-Select} in Thm.~\ref{theorem:det_select_alpha}.

\begin{lemma}
\label{lem:analysis_1}
If $\alpha_u \leq n_u$, then $e_u \in \alpha_u\text{-}rank_{X_u^0}(k_u)$.
\end{lemma}

\begin{pf}
The proof is by induction on $n_u$. The base case is defined to be where $n_u$ is $1$. In this case, $p = k$, and the claim is correct. For the induction step, note that each corruption of an element in $X_u$ can result in at most one rank error. In contrast, corruptions in auxiliary information can result in more than one rank error per corruption, but this is taken care of, as shown next.

By Lemma~\ref{lem:recursion_lemma}, the recursion implementation guarantees that if $\alpha_u \leq n_u$, then $k_u$ is correct. If the return statement in line $13$ is used, then the pivot $x_p$ is returned as $e_u$. The test at line $12$ guarantees that $rank^c_{X_u}(e_u) = k_u$. From the definition of the resilient partition procedure, $e_u \in \alpha_u^{local}\text{-}rank_{X_u^0}(k_u) \subseteq \alpha_u\text{-}rank_{X_u^0}(k_u)$, as needed.

If the return statement in line $13$ is not used, then the return statement in line $19$ is used. Let the last child of $u$ be denoted by $v = v_{|V|}$. The test at line $18$ guarantees that $rank^c_{X_u}(e_u) \in [k_u \pm n_v]$. Notice that the element $e_u$ is the element located by $v$. Therefore, from the definition of the resilient ranking procedure, $e_u \in (\alpha_u^{local} + n_v)\text{-}rank_{X_u^0}(k_u)$. If $\alpha_v > n_v$, then $e_u \in (\alpha_u^{local} + n_v)\text{-}rank_{X_u^0}(k_u) \subseteq (\alpha_u^{local} + \alpha_v)\text{-}rank_{X_u^0}(k_u) \subseteq \alpha_u\text{-}rank_{X_u^0}(k_u)$, as needed.

Otherwise (i.e., if $\alpha_v \leq n_v$), then by induction, $e_u \in \alpha_v\text{-}rank_{X_v^0}(k_v)$. Also, the recursion implementation guarantees that $lb_v$, $ub_v$, and $k_v$, are resilient in this case. If $rank^c_{X_v}(x_p) > k_v$, then $k_v = k_u$, because both are resilient. Also, being that $ub_v$ is resilient, $e_u$ is smaller than all the uncorrupted elements in $X_u$, which are larger than $ub_v$. Therefore, $e_u \in (\alpha_u^{local} + \alpha_v)\text{-}rank_{X_u^0}(k_u)$, as needed. If $rank^c_{X_v}(x_p) < k_v$, then $k_v = k_u - f_u$, because both are resilient. Also, being that $lb_v$ is resilient, $e_u$ is larger than all the uncorrupted elements in $X_u$, which are smaller than $lb_v$. Therefore, $e_u \in (\alpha_u^{local} + \alpha_v)\text{-}rank_{X_u^0}(k_u)$, as needed.
\end{pf}

\begin{lemma}
\label{lem:analysis_2}
Let $v_i = w$ be a first type child of $u$. If $\alpha_w \leq n_w$, then $3n_u/10-3(\alpha_w+\alpha_u^w)-6 \leq rank_{X_u^w}(x_p) \leq 7n_u/10+3(\alpha_w+\alpha_u^w)+6$, where $x_p = e_w$ is the element returned from $w$ to $u$.
\end{lemma}

\begin{pf}
$\alpha_w \leq n_w$, therefore from Lemma~\ref{lem:analysis_1}, it follows that $e_w \in \alpha_w\text{-}rank_{X_w^0}(k_w)$. Also, the recursion implementation guarantees that $k_w$ is resilient in this case, therefore, $k_w = \lceil n_u/10 \rceil$. There exists at least $3(k_w - \alpha_w - 2) - \alpha_u^w$ elements in $X_u^w$ which are smaller than $x_p$. This is because each non corrupted median of five consecutive elements corresponds to at least $3$ elements in $X_u^w$ which are smaller than $x_p$, and each corrupted element either in $X_w$ or in $X_u$ which is not a median of five consecutive elements can change the rank of $x_p$ by at most $1$. A similar argument establishes the second inequality.
\end{pf}

\begin{lemma}
\label{lem:first_phase_repetitions}
Let $w = v_i$ be a first type child of $u$. If $v_{i+1}$ is not a second type node, then $\alpha_u^w + \alpha_w \geq \Omega(n_u)$.
\end{lemma}

\begin{pf}
Being that $w$ is not followed by a second type node, $x_p$ did not pass the test at line $10$ (i.e., $p \notin [f_u, n_u-f_u]$). There are two cases to consider.

If $\alpha_w > n_w = \lceil n_u / 5 \rceil$, then, in particular, $\alpha_w = \Omega(n_u)$.

Otherwise, assume that $\alpha_w < n_w$. It will be shown that $(\alpha_u^w + \alpha_w) \geq n_u/33 = \Omega(n_u)$. Assume, in contradiction, that this is not the case. It follows, from Lemma~\ref{lem:analysis_2}, that $rank^c_{X_u^w}(k_u) \in [3n_u/10-3(\alpha_u^w + \alpha_w)-6, 7n_u/10+3(\alpha_u^w + \alpha_w)+6] \subseteq [3n_u/10-n_u/11-6,7n_u/10+n_u/11+6]$. This contradicts the assumption that $x_p$ did not pass the test at line $10$ (i.e., that $rank^c_{X_u}(k_u) \notin[3n_u/10 - n_u/11 - 6, 7n_u + n_u/11 + 6]$).~\end{pf}

\begin{lemma}
\label{lem:second_phase_repetitions}
Let $w = v_i$ be a second type child of $u$. If $w$ is not the last child of $u$, then $\alpha_u^w + \alpha_w \geq \Omega(n_u)$.
\end{lemma}

\begin{pf}
Being that $w$ is not the last child of $u$, $e_w$ did not pass the test at line $18$ (i.e., $rank_{X_u}(e_w) \notin [k \pm n_w]$). Again, there are two cases to consider.

If $\alpha_w > n_w = n_u - f_u$, then, in particular, $\alpha_w = \Omega(n_u)$.

Otherwise, if $\alpha_w \leq n_w$, then, by Lemma~\ref{lem:analysis_1}, $e_w \in \alpha_w\text{-}rank_{X_w^0}(k_w)$. Moreover, $k_w = k_u$, and each corruption in $X_u$ can cause the rank of $e_w$ to change by at most $1$. Therefore, $rank_{X_u}(e_w) \in [k_u \pm (\alpha_u^w + \alpha_w)]$. However, $e_w$ did not pass the test at line $18$, therefore $\alpha_u^w + \alpha_u > n_w = \Omega(n_u)$.
\end{pf}

\begin{theorem}\label{theorem:det_select_alpha}
\emph{Deterministic-Select} is a deterministic resilient selection algorithm with time complexity $O(n + \alpha)$.
\end{theorem}

\begin{pf}
First, \emph{Deterministic-Select} is shown to be resilient. Let $u$ be the root of the recursion tree, $T$. If $\delta \leq n = n_u$, then by Lemma~\ref{lem:analysis_1}, $e \in \alpha\text{-}rank_{X^0}(k)$, as needed. Otherwise, if $\delta \geq n$, then there are two cases to consider. If $\alpha \leq n$, then by Lemma~\ref{lem:analysis_1}, $e \in \alpha\text{-}rank_{X^0}(k)$, as before. Otherwise, if $\alpha \geq n$, then by definition, $[-\infty,\infty] = n\text{-}rank_{X^0}(k) = \alpha\text{-}rank_{X^0}(k)$. Therefore, for any element $e$, $e \in \alpha\text{-}rank_{X^0}(k)$. In particular, the element returned is correct.

With regard to the time complexity, consider a non-faulty execution (i.e., $\alpha = 0$). The time complexity $T(n) = T(\lceil n/5 \rceil) + T(\lceil 7n/10 \rceil + \lceil n/11 \rceil + 6) + O(n) = O(n)$ follows, because $\lceil n/5 \rceil + \lceil 7n/10 \rceil + \lceil n/11 \rceil < n$.

If $\alpha > 0$, then there might be some repetitions. Lemma~\ref{lem:first_phase_repetitions} and Lemma~\ref{lem:second_phase_repetitions} show that enough corruptions can be charged for the time spent in those repetitions. In particular, the $\Omega(n_u)$ corruptions that cause a first type child repetition pay for the $O(n_u)$ computation time of the child, and the $\Omega(n_u)$ corruptions that cause a second type child repetition pay for the $O(n_u)$ computation time of the child, and for the $O(n_u)$ computation time of the first type child that precedes it. Figure~\ref{fig:recursion_tree} shows a visualization of this amortization. In both cases there is $O(1)$ amortized cost per corruption. Therefore, the overall time complexity is $O(n+\alpha)$.
\end{pf}

\begin{theorem}\label{theorem:det_select}
There exists a deterministic resilient selection algorithm with time complexity $O(n)$.
\end{theorem}

\begin{pf}
The algorithm \emph{Deterministic-Select} can be modified to achieve worst-case time complexity $O(n)$. The algorithm maintains a counter $c$, initialized to $0$, which is a lower bound on the number of corruptions that occurred. Notice that $c$ can be maintained in a reliable memory cell.

The proof of Lemma~\ref{lem:first_phase_repetitions} shows that if a first phase repetition occurred, it must be due to at least $\lfloor n_u / 33 \rfloor$ corruptions, where $u$ is the current node. Therefore, in this case, the counter is incremented by $\lfloor n_u / 33 \rfloor$. The proof of Lemma~\ref{lem:second_phase_repetitions} shows that if a second phase repetition occurred, it must be due to at least $n_v$ corruptions, where $v$ is the second type child of the current node that caused the repetition. Therefore, in this case, the counter is incremented by $n_v$. If the counter is equal to or larger than $n$, the algorithm halts with an arbitrary element.

\newpage
The modified algorithm is correct, because the counter is a lower bound of the number of corruptions. If $c > n$, then $\alpha > n$. Therefore, any element is in the $\alpha$-rank of $X^0$. With regard to the time complexity, notice that the counter is also an upper bound, up to a multiplicative constant, for the amount of extra work performed due to corruptions. Therefore, as long as $c < 2n$, which is always the case, the total work is $O(n)$.
\end{pf}

\section{Recursion Implementation}
\label{sec:recursion_implementation}

In this section, an abstract recursion stack for \emph{Deterministic-Select} is developed. The data structures used by this abstract stack are described, followed by the implementation of the operations on it. This leads to the proof of Lemma~\ref{lem:recursion_lemma} at the end of this section.

\subsection{Data Structures}
\label{subsec:recursion_data_structures}

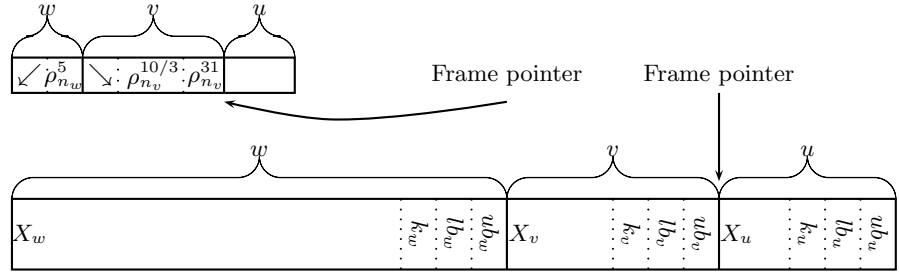
\begin{figure}
\begin{center}
\psset{gridcolor=red,subgridcolor=blue}

\psset{unit=0.47cm}

\begin{pspicture}(26,9)

\pspolygon(1,7)(9,7)(9,8)(1,8)
\psline(3,7)(3,8)
\psline(7,7)(7,8)

\psbrace[rot=270,ref=b,braceWidth=0.01](3,8)(1,8){$w$}
\psbrace[rot=270,ref=b,braceWidth=0.01](7,8)(3,8){$v$}
\psbrace[rot=270,ref=b,braceWidth=0.01](9,8)(7,8){$u$}

\rput(1.5,7.5){$\swarrow$}
\psline[linestyle=dotted](2,7)(2,8)
\rput(2.5,7.5){$\rho_{n_w}^5$}

\rput(3.5,7.5){$\searrow$}
\psline[linestyle=dotted](4,7)(4,8)
\rput(5,7.5){$\rho_{n_v}^{10/3}$}
\psline[linestyle=dotted](5.85,7)(5.85,8)
\rput(6.5,7.5){$\rho_{n_v}^{31}$}

\pscurve{->}(15,6.75)(10,6.25)(7,6.75)
\rput(15,7.5){Frame pointer}

\pspolygon(1,2)(26,2)(26,4)(1,4)
\psline(15,2)(15,4)
\psline(21,2)(21,4)

\psbrace[rot=270,ref=b,braceWidth=0.01](15,4)(1,4){$w$}
\psbrace[rot=270,ref=b,braceWidth=0.01](21,4)(15,4){$v$}
\psbrace[rot=270,ref=b,braceWidth=0.01](26,4)(21,4){$u$}

\psline[linestyle=dotted](12,2)(12,4)
\psline[linestyle=dotted](13,2)(13,4)
\psline[linestyle=dotted](14,2)(14,4)

\psline[linestyle=dotted](18,2)(18,4)
\psline[linestyle=dotted](19,2)(19,4)
\psline[linestyle=dotted](20,2)(20,4)

\psline[linestyle=dotted](23,2)(23,4)
\psline[linestyle=dotted](24,2)(24,4)
\psline[linestyle=dotted](25,2)(25,4)

\rput(1.5,3){$X_w$}
\rput{270}(12.5,3){$k_w$}
\rput{270}(13.5,3){$lb_w$}
\rput{270}(14.5,3){$ub_w$}

\rput(15.5,3){$X_v$}
\rput{270}(18.5,3){$k_v$}
\rput{270}(19.5,3){$lb_v$}
\rput{270}(20.5,3){$ub_v$}

\psline{->}(21,7)(21,4.5)
\rput(21,7.5){Frame pointer}

\rput(21.5,3){$X_u$}
\rput{270}(23.5,3){$k_u$}
\rput{270}(24.5,3){$lb_u$}
\rput{270}(25.5,3){$ub_u$}

\end{pspicture}
\end{center}

\caption{The stacks used by the recursion implementation are depicted. The reliable stack is at the top and the faulty stack is at the bottom. The execution path is composed of the root, $w$, it's first type child, $v$, and $v$'s second type child, $u$. The figure shows the situation when $u$ begins its computation. For each node, the reliable stack stores the traversal direction to its child (drawn as a pointed arrow), as well as the remainders, $\rho^5$ or $\rho^{10/3}$ and $\rho^{31}$, while the faulty stack stores the sub-array $X$, as well as $2g+1$ copies of $lb$, $ub$, and $k$. The frame pointers are also shown.}

\label{fig:recursion_stacks}

\begin{center}
%\line(1,0){440}
\end{center}

\vspace{-0.5in}
\end{figure}

Two stacks, one reliable and the other one faulty, together with a constant number of reliable memory cells, are used to implement the recursion for the algorithm \emph{Deterministic-Select}. An execution path in the recursion tree, $T$, starts from the root and ends at the current node. In each stack, the entire execution path is stored in a contiguous region in memory, where the root is at the beginning, and the current node is at the end. The stacks are depicted schematically in Fig.~\ref{fig:recursion_stacks}.

\subsubsection{Reliable Stack}

The reliable stack stores only $9$ bits of information per node. The height of $T$ is $O(\log n)$, therefore it can be stored in a constant number of reliable memory cells. For each inner node $u \in T$, the reliable stack stores $1$ bit to distinguish between a first type child and a second type child. Let $\rho_x^y$ denote the remainder of the division of $x$ by $y$. For a node of the first type, $\rho^5_{n_u}$ is stored. For a node of the second type, $\rho^{10/3}_{n_u}$ and $\rho^{11}_{n_u}$ are stored. Notice that the $O(1)$ reliable memory cells are used down to the bit level.

\subsubsection{Faulty Stack}

The faulty stack stores $O(n_u)$ words of information per node. For each node $u \in T$, the faulty stack stores the elements of $X_u$, as well as $k_u$, $lb_u$, and $ub_u$. The elements of $X_u$ are stored using $1$ copy per element, while $k_u$, $lb_u$, and $ub_u$ are stored using $2n_u + 1$ copies per variable.

\subsubsection{Global Variables}

Each one of the following global variables is stored using a reliable memory cell:

\begin{itemize}
	\item The current array size
	\item The reliable stack's frame pointer
	\item The faulty stack's frame pointer
	\item The program counter
	\item The return value
\end{itemize}

Notice that at a given moment in an execution only one value per each global variable needs to be stored.

\vspace{0.4in}

%\begin{comment}
%\end{comment}

\subsection{Operations}

Two operations are implemented by the recursion implementation. A push operation corresponds to a recursive call, and a pop operation corresponds to returning from a recursive call.

\subsubsection{Push}

When a node $u$ calls its child $v$, the following is done. The information of whether $v$ is a first type child or a second type child of $u$ is written to the reliable stack, as well as the relevant remainders (i.e., $\rho^5_{n_u}$ or $\rho^{10/3}_{n_u}$ and $\rho^{11}_{n_u}$), and the reliable stack's frame pointer is incremented by $9$ bits. Then, the relevant sub-array is pushed to the faulty stack, followed by the values $lb_v$, $ub_v$, and $k_v$. If $v$ is a first type child, then $n_v$ is updated to $\lceil n_u/5 \rceil$. If $v$ is a second type child, then $n_v$ is updated to $n_u - f_u$. The faulty stack's frame pointer is updated accordingly, and the program counter is set to line $1$. Then, the computation continues to $v$.

\subsubsection{Pop}

When $v$ finishes its computation, the following is done. First, the reliable stack's frame pointer is decremented by $9$ bits, and the information of whether $v$ is a first type or a second type child of $u$ is read, as well as the remainder (i.e., $\rho^5_{n_u}$ or $\rho^{10/3}_{n_u}$ and $\rho^{11}_{n_u}$).

If $v$ is a first type child, then $n_u$ is updated to $5(n_v - 1) + \rho^5_{n_u}$. If $v$ is a second type child, then $n_u$ is updated to $(110/87) \cdot (n_v - \rho^{10/3}_{n_u}/(10/3) + \rho^{11}_{n_u} / 11 - 6)$. Notice that this function is the inverse function of $n_u - f_u$, which is the function used to update $n$ when calling a second type child, as explained before. The faulty stack's frame pointer is decremented by $n_u + 3(2n_u + 1)$ words.

The $2n_u + 1$ copies of $lb_u$, $ub_u$, and $k_u$ are read, and the computed majority of their copies are stored in reliable memory and used as the values for $lb_u$, $ub_u$, and $k_u$. Then, the computation returns to $u$, either to line $8$ or to line $18$, depending on the type of $u$.

\subsection{Proof of Lemma~\ref{lem:recursion_lemma}}

\begin{pf}
The frame pointers, the return value, and the program counter were shown to be reliable, as well as the location of the array $X_u$ and its size $n_u$. $lb_u$, $ub_u$, and $k_u$ are stored using $2n_u + 1$ copies each, therefore, if $\alpha_u \leq n_u$, then these parameters are reliable. The time overhead induced by the frame pointers, return value, program counter, location of the array $X_u$ and its size $n_u$ is a constant. The time overhead induced by $lb_u$, $ub_u$, and $k_u$ is $O(n_u)$. Therefore, the time overhead of the recursive implementation is $O(n_u)$.
\end{pf}

\section{Resilient $k$-d Trees}
\label{sec:resilient_kd_trees}

Gieseke et al.~\cite{kd_trees}, developed a resilient $k$-d tree, where $k$ denotes the dimension (this $k$ is not related to the $k$ in the selection algorithm). As is the case with non-resilient $k$-d trees,
the construction involves multiple partitioning of the points by the median.
For example, if $k = 2$, then at even-depth nodes, the points are partitioned by the $x$-coordinate median,
and at odd-depth nodes, the points are partitioned by the $y$-coordinate median.
In a resilient $k$-d tree, the partitioning ends at the leaves, which contain $b \delta = O(\delta)$ points each,
where $b$ is a parameter.

Gieseke et al. developed a randomized resilient selection algorithm, which is somewhat different
from the randomized resilient selection algorithm developed in this work.
Both algorithms achieve the same expected time complexity, $O(n + \alpha)$.
Using these algorithms to build a resilient $k$-d tree results in $O(n \log n + \delta)$ expected time complexity.
However, the selection algorithm developed here guarantees that the element returned has rank
between $k - \alpha$ and $k + \alpha$ in the input array,
while the algorithm developed in~\cite{kd_trees} only guarantees a rank between $k - O(\delta)$ and $k + O(\delta)$.
This difference does not have asymptotical consequences on the height of the resulting $k$-d tree.

For a deterministic $k$-d tree construction algorithm, Gieseke et al. used the resilient sorting algorithm
developed by Finocchi et al.~\cite{resilient_sorting} in order to partition the points around the median.
This results in $O(n \log^2 n + \alpha\delta)$ time complexity.
By using the deterministic resilient selection algorithm developed here,
the time complexity is reduced to $O(n \log n)$ and implies the following theorem.

\begin{theorem}\label{theorem:kd_trees}
There exists a resilient $k$-d tree which can be constructed in deterministic $O(n \log n)$ time.
It supports resilient orthogonal range queries in $O(\sqrt{n \delta} + t)$ time for reporting $t$ points.
\end{theorem}

\section{Resilient Quicksort Algorithms}
\label{sec:resilient_quicksort_algorithms}

The famous quicksort algorithm is based on the idea of selecting a pivot, partitioning the input by it, and recursively sorting each side of the partition. In the FRAM model the difficulty is in having to maintain the $\omega(1)$ partitioning locations. This is true for both a recursive and iterative implementation. One natural idea for dealing with this difficulty is to partition the array at the median. For sake of simplicity assume that the size of the input is a power of two\footnote{If this is not the case then careful padding can take place. Being that the interest here is in an in-place algorithm, the padding can be done abstractly by knowing that each access to an array location which does not exist can be considered as $\infty$.}. However, using a resilient selection algorithm for locating the median in the FRAM model and partitioning around the element returned does not guarantee that the array is split into two parts of equal size, due to corruptions that may occur during the execution of the selection algorithm, returning an element which is only roughly the median. Thus, there is a need to develop a resilient splitting algorithm, which is defined as follows.

\begin{definition}
A \emph{resilient splitting algorithm} is an algorithm that is given an array $X$ of size $n$ and an integer $k$, and reorders the array such that any uncorrupted element in $X[1, k]$ is smaller than any uncorrupted element in $X[k, n]$.
\end{definition}

In section~\ref{sandboxed_splitting_algorithms}, two non-efficient resilient splitting algorithms are shown: one is deterministic and runs in $O(\alpha n)$ worst-case time, and the second is randomized and in-place and runs in $O(\alpha n)$ expected time.

In section~\ref{generic_splitting_algorithms}, two efficient resilient splitting algorithms are shown: one is deterministic and runs in $O(n + \alpha\delta)$ worst-case time, and the second is randomized and in-place and runs in $O(n + \alpha\delta)$ expected time. These efficient algorithms use the non-efficient algorithms from Section~\ref{sandboxed_splitting_algorithms}.

\subsection{Sandboxed Splitting Algorithms}
\label{sandboxed_splitting_algorithms}

The basic idea behind the resilient splitting algorithms is to test the rank of the element returned by the selection algorithm, and \emph{fix} it, as needed. In order to achieve this goal, the notion of \emph{Sandboxing} an algorithm is introduced. The idea is to convert a non-resilient algorithm $A$, with a known bound on its running time and space usage into a resilient algorithm $A'$. However, in order to be able to do this, there must exist a verification procedure which can verify that the output of $A$ is correct, and the algorithm $A$ needs to be \emph{non-destructive}, a notion which is defined later.

Finocchi, Grandoni and Italiano~(\cite{resilient_dictionaries}, Lemma 4) already considered a similar reduction. However, there is an unfortunate flaw in their proof given there, because it does not take into consideration the following two cases: A corrupted variable that can cause the non-resilient procedure to require a much larger time complexity (maybe even getting stuck in an infinite loop), and memory corruptions that can cause the non-resilient procedure to alter memory cells used by other parts of the system. These problems can be overcome by confining the execution to a predetermined area in memory and having an upper bound on the running time of $A$. The area in memory is referred to as the \emph{sandbox}. For a problem $P$ and an input $X$, let $P(X)$ denote the set of correct solutions of $P$ on $X$.

\begin{definition}
Let $A$ be an algorithm for a problem $P$. Assume that an execution of $A$ on an input $X$ can be interrupted at any point in time, and let $X'$ denote the state of the input after such an interruption. $A$ is \emph{non-destructive} if for any execution of $A$ on any input $X$ and on any set of random coins, and for any interruption with any possible sequence of faults, $P(X) = P(X')$.
\end{definition}

\begin{lemma}
\label{lem:sandbox_lemma}
Let $A$ be a non-resilient and a non-destructive algorithm solving problem $P$ with time complexity $T_A$ (either worst-case or expected) and space complexity $S_A$. Let $C$ be a resilient verification procedure for $A$ with worst-case time complexity $T_C$ and space complexity $S_C$ which decides the correctness of an execution of $A$.
Then there exists a resilient algorithm $A'$ which solves $P$, and has time complexity $O((1 + \alpha)(T_A + T_C))$ which is either worst-case or expected, depending on $T_A$, and space complexity $O(S_A + S_C)$.
\end{lemma}

\begin{pf}
A \emph{sandboxed} version of $A$, denoted by $A'$, is defined as follows. The algorithm works in rounds. In each round, $A'$ runs a modified version of $A$, as defined next. If the running time of $A$ is worst-case, then, to guarantee that $A$ will not run for too long, $A'$ runs $A$ for no more than $T_A$ steps. If the running time of $A$ is in expectation, then, to guarantee that $A$ will not run for too long, $A'$ runs $A$ for no more than $2T_A$ steps. This is done by counting the number of computational steps that $A$ performs. To guarantee that $A$ will not alter memory cells other than its own, $A'$ runs $A$ confined to a memory region of size $S_A$. The counter which counts the computational steps as well as the two boundaries for the memory region are stored in reliable memory cells. After running the modified $A$ algorithm, $A'$ calls $C$ to check the correctness of $A$'s computation. If $C$ returned a positive answer, $A'$ halts. Otherwise, a new round begins, but only after the memory sandbox is flushed. That is, immediately after a non-successful round ends, all of the working memory is erased, but the input is left as it is, for the next round.

The memory sandbox guarantee that the (non-resilient) computation of $A$ would not alter memory cells outside of $A$'s computation. $A'$ halts only after the resilient verification procedure $C$ returned a positive answer. Therefore, $A'$ is correct, even in the presence of memory faults.

If the running time of $A$ is worst-case, then each round takes $T_A+T_C$ time. In a non-faulty round, $A$ is correct. By the pigeon hole principle, if there are more than $\alpha$ rounds, at least one of them is non-faulty. Denote the state of the input at the beginning of this non-faulty round by $X'$. $A$ is a (non-resilient) algorithm for $P$, therefore, in this non-faulty round, it computes a correct output $y \in P(X')$. $A$ is non-destructive, therefore $P(X) = P(X')$. It follows that $y \in P(X)$, i.e., $y$ is correct\footnote{If $A$ is a Monte Carlo algorithm then $\mathbb{P}\left[y \in P(X')\right] \geq \frac{2}{3}$, and because of the non-destructiveness of $A$, it follows that $\mathbb{P}\left[y \in P(X)\right] \geq \frac{2}{3}$, as needed.}. Therefore, there are at most $\alpha+1$ rounds.

If the running time of $A$ is expected, then each round takes $2T_A+T_C$ time. In a non-faulty round, the probability that $A$ would halt within $2T_A$ computational steps is at least $\frac{1}{2}$, by Markov's inequality. Therefore, the expected number of rounds is at most $2 \alpha + 1$.

The space used by the calls to $A$ and $C$ can be reused, therefore, the space complexity is $O(S_A + S_C)$.~\end{pf}

This general notion of sandboxed algorithms can be used for designing resilient splitting algorithms. For the deterministic resilient splitting algorithm, algorithm $A$ is executed by using the non-resilient deterministic selection algorithm to locate the median, and partitioning the array around the element returned. The verification procedure $C$ is implemented by testing that each side of the partition has the same size. For the randomized resilient splitting algorithm, algorithm $A$ is executed by using the non-resilient randomized selection algorithm to locate the median. Notice that in the randomized case, $S_A = O(1)$.

Notice that both the non-resilient deterministic selection algorithm and the non-resilient randomized selection algorithm needs to slightly be altered in order to be non-destructive. The only operation that these algorithms perform which might alter the input is swapping. The idea is to make these swaps atomic, i.e., only stopping the algorithm after such a swap is fully completed. Notice that the $k$-order statistic of an input array does not depend on the specific permutation of the input array\footnote{Another way of altering these algorithms to make them non-destructive is by copying the input array to a second and temporary array. Then, performing all of the swaps only on the temporary array, and making sure that the input array is not altered at all, by putting it outside the memory sandbox. This solution, however, has a cost in time and space.}.

\begin{corollary}\label{lem:sandbox_split}
There exists a deterministic resilient splitting algorithm with worst-case time complexity $O(\alpha n)$, and a randomized in-place resilient splitting algorithm with expected time complexity $O(\alpha n)$.
\end{corollary}

\begin{pf}
The proof follows from Lemma~\ref{lem:sandbox_lemma} and from the discussion above.
\end{pf}

Denote the algorithm from Lemma~\ref{lem:sandbox_split} by \emph{Sandboxed-Split}. The running time of such an algorithm is rather costly, but it is still useful when considering small arrays. For the resilient splitting algorithm, the idea is to reduce the size of the array, and then execute Sandboxed-Split.

\subsection{Efficient Splitting Algorithms}
\label{generic_splitting_algorithms}

Consider the following generic algorithm, denoted by \emph{Generic-Resilient-Split}. The algorithm uses either \emph{Deterministic-Select} or \emph{Randomized-Select}, denoted here by \emph{Generic-Resilient-Select}, to locate both the $(k-\delta)^{th}$ and the $(k+\delta)^{th}$ order statistics. Then, it uses \emph{Sandboxed-Split} to split the remaining $O(\delta)$ elements.\\ \ \\

\begin{algorithm}[H]

$l \leftarrow$ Generic-Resilient-Select($X[1, n]$, $k - \delta$)\;
partition $X$ around $l$\;
\If{$rank^c_X(l) = k$}{\Return}

$r \leftarrow$ Generic-Resilient-Select($X[l, n]$, $k - l + \delta + 1$)\;
partition $X[l, n]$ around $r$\;
\If{$rank^c_X(r) = k$}{\Return}

Sandboxed-Split($X[l, r]$, $k - l$)\;

\caption{Generic-Resilient-Split($X$, $k$)}
\end{algorithm}
\vspace{0.4cm}

\begin{lemma}
\label{lem:generic_resilient_split_lemma}
\emph{Generic-Resilient-Split} is a resilient splitting algorithm.
\end{lemma}

\begin{pf}
After the array is partitioned around $l$, the uncorrupted elements in $X[1, l]$ are smaller than the uncorrupted elements in $X[l, n]$. After the array is partitioned around $r$, the uncorrupted elements in $X[l, r]$ are smaller than the uncorrupted elements in $X[r, n]$. After the call to \emph{Sandboxed-Split}, the uncorrupted elements in $X[l, k-l]$ are smaller than the uncorrupted elements in $X[k-l, r]$. It follows that the uncorrupted elements in $X[1, k]$ are smaller than the uncorrupted elements in $X[k, n]$, as needed.~\end{pf}

The following corollaries follow by substituting Generic-Resilient-Select by either the deterministic or randomized versions. Notice that in both cases, the call to \emph{Generic-Resilient-Select} takes $O(n + \alpha)$ (either in expectation or worst-case as needed). The size of the sub-array $X[l, r]$ is $O(\delta)$, therefore the call to \emph{Sandboxed-Select} takes $O(\alpha\delta)$ time.

\begin{corollary}
\label{cor:resilient_split}
There exists a deterministic resilient splitting algorithm with worst-case running time $O(n+\alpha\delta)$, and a randomized in-place resilient splitting algorithm with expected running time $O(n+\alpha\delta)$.
\end{corollary}

\subsection{Resilient Quicksort Algorithms}
Using the generic resilient splitting algorithm as a black box, one can sort resiliently using \emph{Generic-Resilient-Quicksort}. Notice that this algorithm does not use more than $O(1)$ space, except for the space used by the splitting algorithm.\\

\begin{algorithm}[H]
\label{alg:Resil_quicksort}

\For{$d \in [1..\log n]$}{
	\For{$c \in [0..2^d-1]$}{
		$n' \leftarrow n / 2^d$ \;
		$X' \leftarrow X[n'\cdot c + 1, n'\cdot (c+1)]$\;
		\# The array is split in-place. \;
		Generic-Resilient-Split($X'$, $n'/2$)\;
	}
}
\caption{Generic-Resilient-Quicksort($X$)}
\end{algorithm}

\begin{lemma}\label{lem:generic_resil_quicksort}
\emph{Generic-Resilient-Quicksort} is a resilient sorting algorithm.
\end{lemma}

\begin{pf}
Consider two uncorrupted elements $a$ and $b$ from the input, where $a<b$. There exists some element $p$ which partitions them, at which point $a$ will be put before $b$ in the array, and from then onwards their order will remain the same.
\end{pf}

The following theorem follows.

\begin{theorem}
There exists a deterministic sorting algorithm with worst-case running time of $O(n\log n+\alpha\delta)$, and a resilient randomized in-place sorting algorithm with expected running time of $O(n\log n+\alpha\delta)$.
\end{theorem}

\begin{pf}
Using Corollary~\ref{cor:resilient_split} and Lemma~\ref{lem:generic_resil_quicksort}, the theorem follows.
\end{pf}

\newpage

\bibliographystyle{alpha}
\bibliography{arxiv_bib}

\end{document}